\documentclass[%
 reprint,
 amsmath,amssymb,
 aps,
]{revtex4-2}

\usepackage{textgreek} % for Greek letters without math-mode
\usepackage{upgreek}
\usepackage{lipsum}     % for dummy text
\usepackage[dvipsnames]{xcolor} % for colour text % https://en.wikibooks.org/wiki/LaTeX/Colors
\usepackage{gensymb}

\usepackage{graphicx}% Include figure files
\usepackage{dcolumn}% Align table columns on decimal point
\usepackage{bm}% bold math
\begin{document}

\preprint{APS/123-QED}

%\title{Manuscript Title:\\with Forced Linebreak}% Force line breaks with \\
\title{The Effect of Ultrastrong Magnetic Fields on Laser-Produced Gamma-Ray Flashes}

\author{P. Hadjisolomou}
\email{Prokopis.Hadjisolomou@eli-beams.eu}
\author{R. Shaisultanov}
\author{T. M. Jeong}
\author{P. Valenta}
\author{S. V. Bulanov}
\altaffiliation{National Institutes for Quantum Science and Technology (QST), Kansai Photon Science Institute, 8-1-7 Umemidai, Kizugawa, Kyoto 619-0215, Japan}
\affiliation{
ELI Beamlines Facility, Extreme Light Infrastructure ERIC, Za Radnicí 835, 25241 Dolní Břežany, Czech Republic
}

\date{\today}

\begin{abstract}
Laser produced \textgamma-photons can make an important impact on applied and fundamental physics that require high \textgamma-photon yield and strong collimation. We propose addition of a constant magnetic field to the laser-solid interaction to obtain the aforementioned desired \textgamma-photon properties. The \textgamma-ray flash spatial and spectral characteristics are obtained via quantum electrodynamics particle-in-cell simulations. When the constant magnetic field aligns with the laser magnetic field then the \textgamma-ray emission is significantly enhanced. Moreover, the \textgamma-photon spatial distribution becomes collimated, approximately in the form of a disk.
\end{abstract}

\maketitle

%---------------------------------------------------------------------------------------------------------------------------------------------------------------------------
%---------------------------------------------------------------------------------------------------------------------------------------------------------------------------

%%%%%%%%%%%%%%%%%%%%%%%%%%%%%%%%%%%%%%%%%%%%%%%%%%%%%%%%%%%%%%%%%%%%%%
%%%%%%%%%%%%%%%%%%%%%%%%%%%%%%%%%%%%%%%%%%%%%%%%%%%%%%%%%%%%%%%%%%%%%%

%%%%%%%%%%%%%%%%%%%%%%%%%%%%%%%%%%%%%%%%%%%%%%%%%%%%%%%%%%%%%%%%%%%%%%
%%%%%%%%%%%%%%%%%%%%%%%%%%%%%%%%%%%%%%%%%%%%%%%%%%%%%%%%%%%%%%%%%%%%%%

%{\color{blue}

\par The 1980's witnessed the invention of the Chirped Pulse Amplification technique \cite{1985_StricklandD} resulting in the rapid growth of laser power, exceeding the petawatt (PW) level by the end of the 20th century \cite{1999_Perry}. The $10 \kern0.2em \mathrm{PW}$ lasers \cite{2019_DansonC} became reality with a femtosecond-class laser in ELI-NP, another laser of ten times higher energy being near completion in ELI-Beamlines and the upgrade of the Apollon laser.

\par When a PW-class laser interacts with matter energetic charged particles are generated, being the norm over the past years. Once multi-PW laser facilities came forward, generation and usage of \textgamma-photons became one of the main tasks of those facilities. A high laser to \textgamma-photon energy conversion efficiency, $\kappa_\gamma$, is predicted \cite{2012_NakamuraT, 2012_RidgersCP, 2022_GonoskovA}. The \textgamma-ray flashes suit, among others, photonuclear reactions \cite{2022_KolenatyD}, positron sources \cite{2015_SarriG}, neutron sources \cite{2014_PomerantzI}, extreme energy density materials science \cite{2013_EliassonB} and studies on fundamental processes \cite{2023_MacLeodAJ}.

\par The \textgamma-ray flash is a result of the multiphoton Compton scattering process \cite{1985_RitusVI}, which occurs when an electron collides with the laser, emitting a scattered \textgamma-photon. The scattering process reads $\mathrm{e^-} + N_l \omega_l \rightarrow \mathrm{e^-} + \omega_\gamma$, where $\mathrm{e^-}$ represents an electron, $N_l \gg 1$ is the number of laser photons, $\omega_l$ is the laser frequency, and $\omega_\gamma$ is the scattered \textgamma-photon frequency. The \textgamma-photon yield is quantified by the quantum nonlinearity parameter, $\chi_e =  \gamma_e E_S^{-1} \sqrt{ ({\bf{E}} + {\bf{v}} \times {\bf{B}})^2 - ({\bf{v}} \cdot {\bf{E}} / c)^2 } $. Here, $E_S = m_e^2 c^3 / (e \hbar) \approx 1.3 \times 10^{18} \kern0.2em \mathrm{V m^{-1}}$ is the Schwinger field ($m_e$ is the electron rest mass, $c$ is the vacuum speed of light, $e$ is the elementary charge and $\hbar$ is the reduced Planck constant), $\gamma_e$ is the Lorentz factor of an electron with velocity $\bf{v}$ before scattering, ${\bf{E}}$ is the electric field and ${\bf{B}}$ is the magnetic field. High $\kappa_\gamma$ values require $\chi_e \gg 1$ \cite{2012_NakamuraT, 2012_RidgersCP}.

\begin{figure} % figure figure figure figure figure figure figure figure figure figure figure
  \centering
  \includegraphics[width=0.85\linewidth]{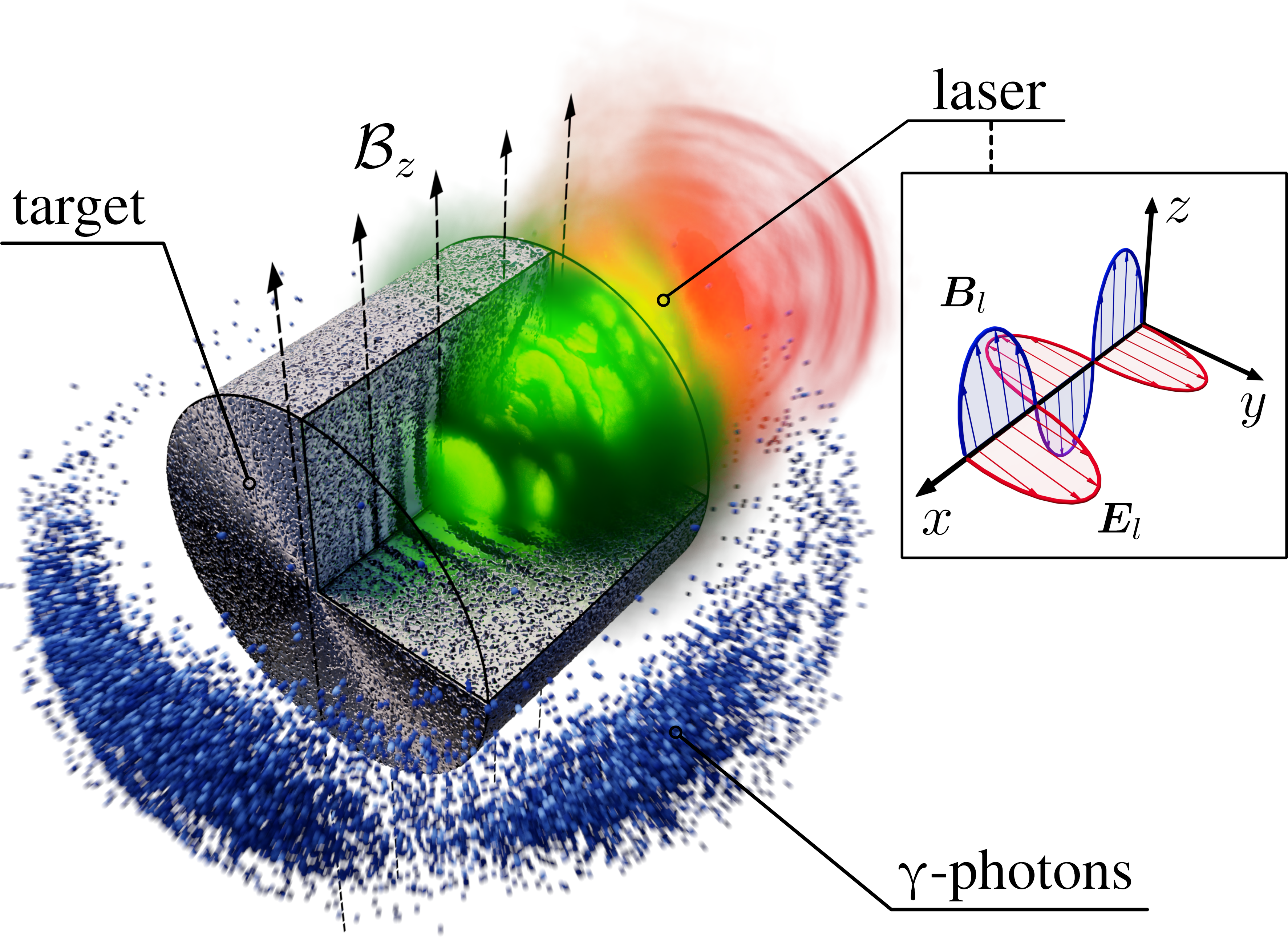}
  \caption{An $a=350$ laser (figure inset) interacting with a lithium foil, in the presence of a CMF (black dashed arrows), at $80 \kern0.2em \mathrm{fs}$ PIC simulation time. The case where the CMF corresponds to the $b_z=150$ case is depicted. The laser-foil interaction results in enhanced \textgamma-photon emission, illustrated by blue spheres (energy larger than $500 \kern0.2em m_e c^2$, with a ballistic offset of $10 \kern0.2em \mathrm{\upmu m}$). The green-red and the gray colors show the laser field and the target respectively.}
  \label{fig:1}
\end{figure} % figure figure figure figure figure figure figure figure figure figure figure

\par The purpose of this paper is to demonstrate that a constant magnetic field (CMF) added to a laser field can enhance the \textgamma-photon yield and directionality from laser-solid interactions. We initially study the single-electron dynamics under the effect of an ultraintense laser in addition to a CMF, with radiation reaction force taken into account. We then extend our study to particle-in-cell (PIC) simulations with the target being a lithium slab, as shown in Fig. \ref{fig:1}. It has recently been shown that lithium combined with multi-PW lasers results in high $\kappa_\gamma$ \cite{2022_HadjisolomouP_b}; although generation of a small electron-positron pair population through the multiphoton Breit-Wheeler process is possible under the conditions examined, it does not affects the \textgamma-photon emission and is therefore ignored. We demonstrate that a strong CMF orthogonal to the laser propagation direction increases $\kappa_\gamma$ by several times. Moreover, if the CMF aligns with the laser magnetic field then the \textgamma-ray flash distribution resembles a narrow disk along the laser electric field oscillating plane, increasing the emitted \textgamma-photon fluence.

%}

%%%%%%%%%%%%%%%%%%%%%%%%%%%%%%%%%%%%%%%%%%%%%%%%%%%%%%%%%%%%%%%%%%%%%%

%{\color{red}

\par Let us measure velocity in $c$, momentum in $m_e c$, distance in $\lambdabar=\lambda/ (2 \pi)$ (where $\lambda = 0.815 \kern0.2em \mathrm{\upmu m}$ is the laser wavelength) and time in $\omega_l^{-1}$ (where $\omega_l=c / \lambdabar$). We use a linearly polarized electromagnetic wave with an electric field $\bm{E}_l=[0, a \cos(t-x), 0]$ and a magnetic field $\bm{B}_l=[0, 0, a c^{-1} \cos(t-x)]$. The normalized dimensionless amplitude of the wave is  $a = e E_l / (m_e c \omega_l) \approx 350$. The normalized dimensionless amplitude of the CMF, $\bm{\mathcal{B}}$, is $\bm{b} = e \bm{\mathcal{B}} / (m_e \omega_l) = [b_x, b_y, b_z]$ and its effect on the electron motion is studied for the values of 75, 150 and 300. These fields are in the megatesla scale for the wavelength under consideration \cite{2002_TatarakisM}.

\par A radiation reaction force, ${\bm{F}}_{rad}= -\bm{p} 2 a m_e c^2 \chi_e^2 G_e /(3 \hbar \gamma_e)$, is considered \cite{1980_LandauLD}, where ${\bm{p}}=[p_x, p_y, p_z]$ is the electron momentum and $G_e=(1+8.93 \chi_e+2.41 \chi_e^2)^{-2/3}$ is the Gaunt factor which accounts for quantum electrodynamics correction of the radiation reaction effect \cite{2022_GonoskovA}. The equations of motion are
\begin{equation}
\frac{d\bm{r}}{dt}=\frac{\bm{p}}{\gamma_e}
\label{eq1}
\end{equation}
and
\begin{equation}
\frac{d\bm{p}}{dt}=-\left[ \bm{E}_l + \frac{\bm{p}}{\gamma_e}\times (c \bm{B}_l+\bm{b}) + \frac{2 \alpha m_e c^2}{3 \hbar \omega_l} \chi_e^2 G_e \frac{\bm{p}}{\gamma_e} \right] .
\label{eq2}
\end{equation}
Here,
$\bm{r}=[x, y, z]$,
$\alpha = e^2 / (4 \pi \varepsilon_0 \hbar c)$ is the fine structure constant and $\varepsilon_0$ is the vacuum permittivity.

%}

%%%%%%%%%%%%%%%%%%%%%%%%%%%%%%%%%%%%%%%%%%%%%%%%%%%%%%%%%%%%%%%%%%%%%%

\begin{figure} % figure figure figure figure figure figure figure figure figure figure figure
  \centering
  \includegraphics[width=1.0\linewidth]{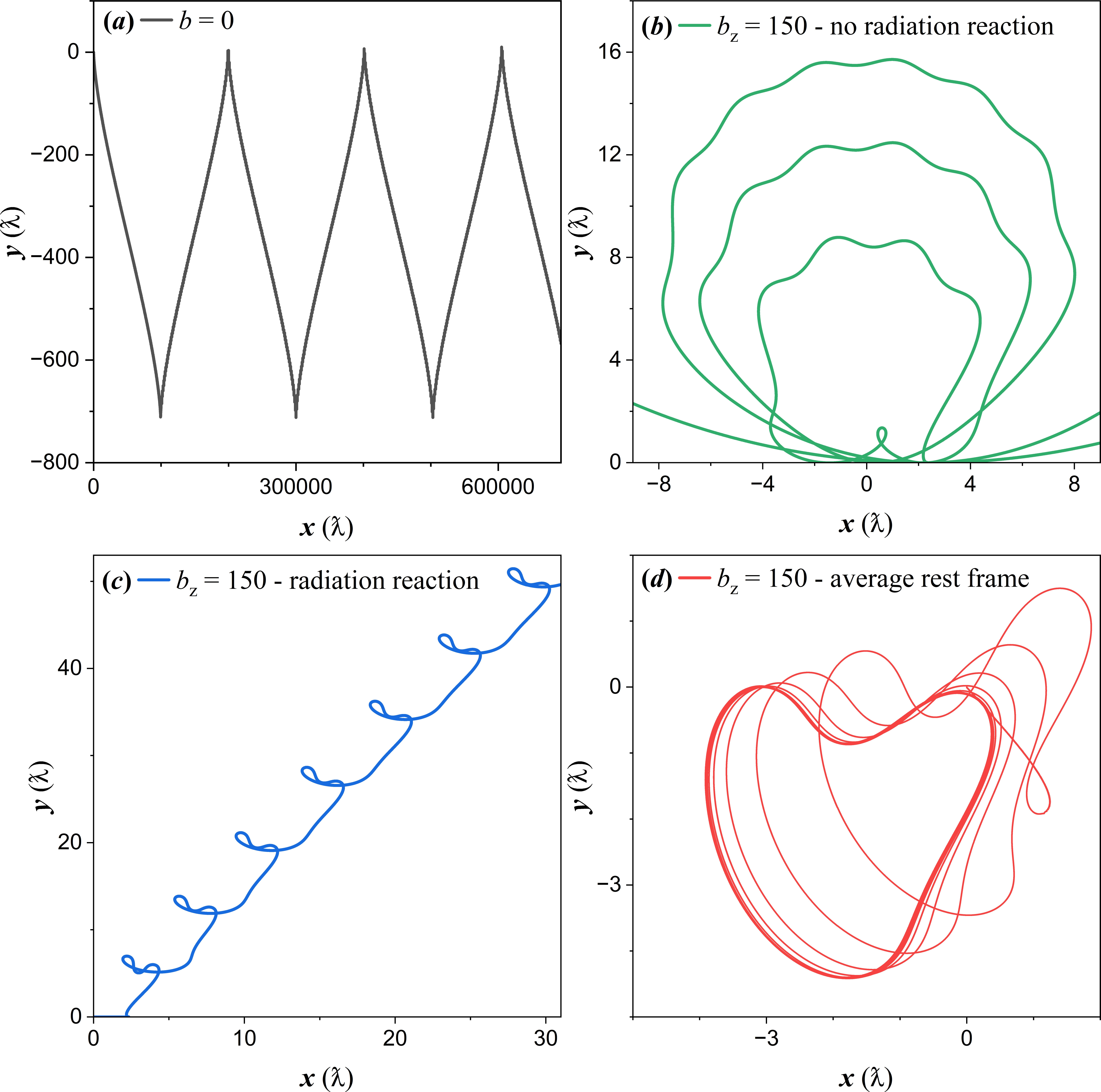}
  \caption{Electron trajectory by solving equations of motion under the influence of a linearly polarized electromagnetic wave of $a \approx 350$ plus a CMF of (a) $b=0$, (b) $b_z=150$ ignoring radiation reaction, (c) $b_z=150$ including radiation reaction and (d) $b_z=150$ including radiation reaction, in electron average rest frame.}
  \label{fig:2}
\end{figure} % figure figure figure figure figure figure figure figure figure figure figure

\par The obtained single-electron trajectories are shown in Fig. \ref{fig:2}(a-d). Fig. \ref{fig:2}(a) corresponds to the reference case where only a plane electromagnetic wave is considered \cite{1971_GunnJE}, without CMF. The electron drifts along the x-axis with a drift velocity, $u_d \lesssim c$; the electron trajectory in the electron average rest frame is a figure-eight. The radiation reaction force is maximized where the trajectory curvature is high. Projection of electron trajectories at those locations indicates that the radiation is emitted at two symmetric lobes \cite{2012_NakamuraT, 2022_HadjisolomouP_b}.

\par Let us set $b=150$, except where otherwise mentioned. For the $b_x$ case (see supplementary figure) the electron trajectory forms periodic spiral-like patterns, with the period decreasing for increasing $b_x$. The ratio of its z-axis to y-axis extent is $b_x$. The radiation reaction force initially obtains low values of $0.0018 m_e c \omega_l$ and slowly decreasing thereafter. The particle trajectory is approximately the same whether or whether not the radiation reaction is considered, with $\gamma \approx 12$. Note that the $b_x$ case combined with a circularly polarized laser in the absence of radiation reaction has been studied six decades ago, referred to as the autoresonance effect where an electron can be accelerated to high energies \cite{1963_KolomenskiiAA_b, 1963_DavydovskiiVY}. However, inclusion of radiation reaction suppresses the autoresonance scheme \cite{1972_ZeldovichYB, 2015_SagarV}.

\par For the $b_y$ case (see supplementary figure) the electron propagation direction rotates approximately $-12 \degree$ (for the parameters defined) on the xz-plane. By shifting the electron trajectory along the new propagation axis the trajectories form figure-eight patterns, with the trajectory period decreasing for increasing $b_y$. Thus, the highest \textgamma-photon energy yield is expected at two lobes located at negative z-values. The Lorentz factor obtains a peak value of $\gamma \approx 390$, whilst the radiation reaction force reaches values of $\approx 55 m_e c \omega_l$, revealing strong \textgamma-photon emission.

\par If the radiation reaction is ignored, the $b_z$ case gives circular trajectories along the xy-plane as shown in Fig. \ref{fig:2}(b), passing through $z=0$ and symmetrically around the y-axis; no drifting of the electron trajectory occurs \cite{2022_BeloborodovAM}. In this case, the Lorentz factor is $\mathcal{O}(10^3)$. The picture changes by the inclusion of radiation reaction, where the electron trajectory drifts along an axis (with $u_d \approx 0.518$) forming an angle $\theta \approx \arccos(1 - b_z/a) \approx 55 \degree$ with the laser propagation axis. The electron trajectories form periodic high-curvature patterns, as shown in Fig. \ref{fig:2}(c). By considering the electron average rest frame then a heart-figure is revealed, as seen in Fig. \ref{fig:2}(d). The Lorentz factor peaks at $\gamma \approx 1440$, with a period matching that of the high-curvature patterns appearance. The $b_z$ case corresponds to the highest radiation reaction force values, at $\approx 320 m_e c \omega_l$, and strongest \textgamma-photon emission is expected for this case. Due to the electron trajectory restricted on the xy-plane, all \textgamma-photons are emitted as a \textgamma-ray disk.

%%%%%%%%%%%%%%%%%%%%%%%%%%%%%%%%%%%%%%%%%%%%%%%%%%%%%%%%%%%%%%%%%%%%%%

%{\color{red}

\par Now we realize the laser-target interaction through three-dimensional relativistic quantum electrodynamic PIC simulations. We use the EPOCH \cite{2015_ArberTD, 2014_RidgersCP} code, enabling the Photons (acknowledging the radiation reaction effect) and the Higuera-Cary (obtaining accurate high-energy electron trajectories) preprocessor directives. For similar interaction parameters, in the absence of the CMF, Compton \textgamma-photons dominate over Bremsstrahlung \textgamma-photons within the simulation time \cite{2022_HadjisolomouP_b}. Therefore, the Bremsstrahlung preprocessor directive is not enabled. Although a small population of electron-positron pairs (orders of magnitude smaller than the target electrons) is generated during the interaction \cite{2022_HadjisolomouP_b}, they are ignored as we are only interested in the \textgamma-ray flash characterization.

\par The laser and CMF parameters used in the PIC simulations match those used in the single-electron model, where now the pulsed nature of the laser is considered with $17 \lambdabar$ focused diameter, $40 \omega_l^{-1}$ pulse duration and a pulse temporal offset of two standard deviations. Three simulation sets are performed, each having the CMF oriented along one of the three Cartesian axes. The laser focuses in the center of the simulation box, with box open boundaries at $\pm 15.36 \kern0.2em \mathrm{\mu m}$ in each direction. Normal laser incidence on the target (yz-plane) requires cells of $10 \kern0.2em \mathrm{nm}$ along the x-axis and $40 \kern0.2em \mathrm{nm}$ on the other two, where eight macroelectrons and eight macroions are assigned per cell. The laser peak reaches the focal spot at $65 \kern0.2em \mathrm{fs}$, while it requires $110 \kern0.2em \mathrm{fs}$ for \textgamma-photon emission to increase asymptotically, indicating that Compton \textgamma-photon emission occurs in timescales comparable to the laser duration.

%}

%%%%%%%%%%%%%%%%%%%%%%%%%%%%%%%%%%%%%%%%%%%%%%%%%%%%%%%%%%%%%%%%%%%%%%

\begin{figure} % figure figure figure figure figure figure figure figure figure figure figure
  \centering
  \includegraphics[width=1.0\linewidth]{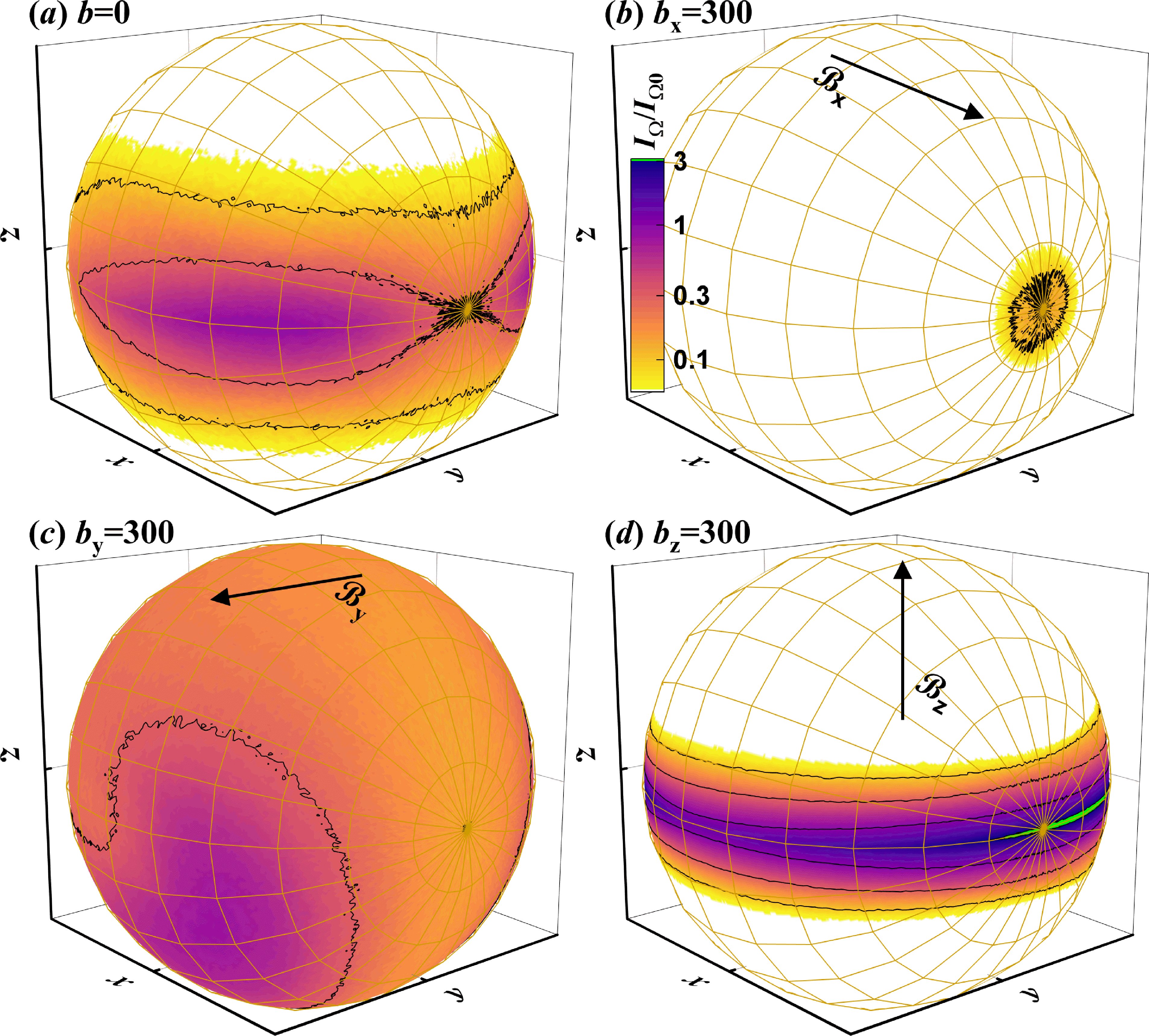}
  \caption{$I_\Omega / I_{\Omega 0}$ as obtained by PIC simulations of an $a=350$ laser interacting with a lithium target in the presence of a CMF, for the cases of (a) $b=0$, (b) $b_x=300$, (c) $b_y=300$ and (d) $b_z=300$. For the $b_z=300$ case the color-bar saturates (green color).}
  \label{fig:3}
\end{figure} % figure figure figure figure figure figure figure figure figure figure figure

\par The \textgamma-photon radiant intensity, $I_\Omega$, being the total \textgamma-photon energy emitted per solid angle per pulse duration maps the \textgamma-photon directionality. The $b=0$ case corresponds to a peak $I_\Omega$ reference value of $I_{\Omega 0}$. Fig. \ref{fig:3}(a-d) depicts $I_\Omega/I_{\Omega 0}$, with the CMF orientation labeled in the figure. Fig. \ref{fig:3}(a) shows a \textgamma-photon double lobe structure, as predicted by previous PIC simulations and in agreement with our single-electron model.

\par Fig. \ref{fig:3}(b) corresponds to the $b_x$ case, where weak radiation reaction effects occur, and peak $I_\Omega$ is more than three times lower than the $b=0$ case. The radially symmetric distribution of the \textgamma-photons geometrically result in a collimated \textgamma-ray flash along the laser propagation axis. The $b_y$ case shown in Fig. \ref{fig:3}(c) has peak $I_\Omega$ comparable (within $10 \kern0.2em \%$) to $I_{\Omega 0}$. However, the lobes are enlarged, revealing a larger $\kappa_\gamma$ value.

\par The \textgamma-ray disk reappears in Fig. \ref{fig:3}(d), corresponding to the $b_z$ case. Here, the \textgamma-photons pile-up on a narrow distribution on the xy-plane. Projection of the electron trajectories (see Fig. \ref{fig:2}(c)) suggests that the \textgamma-ray disk is not uniform along the xy-plane. For the $b_z \approx 300$ case, a lobe of $I_\Omega/I_{\Omega 0}>3$ occurs, squeezed within an angle of $< 10 \degree$ at full-width-at-half-maximum along z-axis.

\begin{figure} % figure figure figure figure figure figure figure figure figure figure figure
  \centering
  \includegraphics[width=0.85\linewidth]{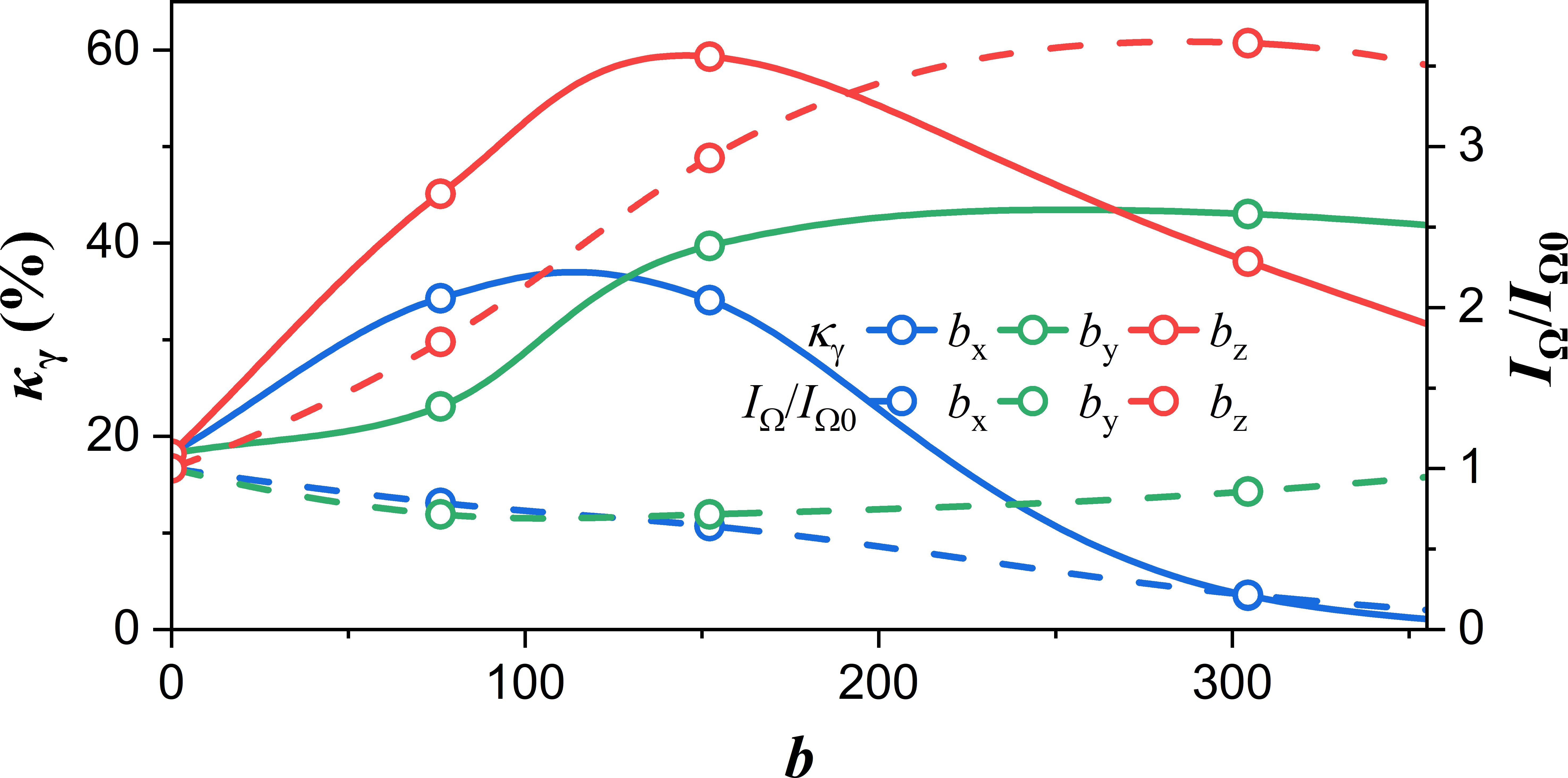}
  \caption{left axis - solid lines: $\kappa_\gamma$ as a function of $b$, due to the interaction of an $a=350$ laser with a lithium foil in the presence of a CMF. $b_x$, $b_y$ and $b_z$ correspond to the blue, green and red lines respectively, reaching a $\kappa_\gamma$ value of $60 \kern0.2em {\%}$ for the $b_z \approx 150$ case. right axis - dashed lines: Peak $I_\Omega/I_{\Omega 0}$ as a function of $b$.}
  \label{fig:4}
\end{figure} % figure figure figure figure figure figure figure figure figure figure figure

\par Our results, expanded over a CMF amplitude range, are summarized in Fig. \ref{fig:4}. The $b_x$, $b_y$ and $b_z$ cases are represented by the blue, green and red lines respectively. The continuous lines show $\kappa_\gamma$ as a function of $b$. If no magnetic field is applied, the laser interaction with the lithium target results in $\kappa_\gamma \approx 20 \kern0.2em \%$. For the $b_x$ case, applying a CMF up to $b_x \approx 150$ increases $\kappa_\gamma$, but further increase of $b_x$ results in rapid decrease of $\kappa_\gamma$. This effect is due to magnetically induced transparency of the target \cite{2005_KryachkoAY}, where a highly transparent target results in weak laser-target interaction, therefore low \textgamma-photon emission. The dashed blue line shows $I_\Omega/I_{\Omega 0}$ steadily decreasing for increasing $b_x$.

\par Our PIC simulations reveal that magnetically induced transparency does not occur if the CMF is along y-axis, within our parameters of interest. Although $I_\Omega/I_{\Omega 0}$ remains approximately constant, the extent of the \textgamma-photon lobes increases. As a result, $\kappa_\gamma$ gradually increases, reaching $\kappa_\gamma \approx 40 \kern0.2em \%$ for $b_y=300$.

\par Target transparency also occurs for the $b_z$ case \cite{1998_TeychenneD, 2003_HurMS, 2021_MandalD}, where $b_z > 150$ results in decrease of $\kappa_\gamma$. However, at  $b_z = 150$ we obtain $\kappa_\gamma \approx 60 \kern0.2em \%$, three times higher than the reference case. The $b_z$ case is of particular interest also due to its high $I_\Omega$, as seen by the red dashed line in Fig. \ref{fig:4}(a). Although optimal $\kappa_\gamma$ occurs at $b_z = 150$, optimal $I_\Omega/I_{\Omega 0}$ occurs at $b_z = 300$ due to a narrower (along z-axis) \textgamma-photon lobe.

\begin{figure} % figure figure figure figure figure figure figure figure figure figure figure
  \centering
  \includegraphics[width=0.8\linewidth]{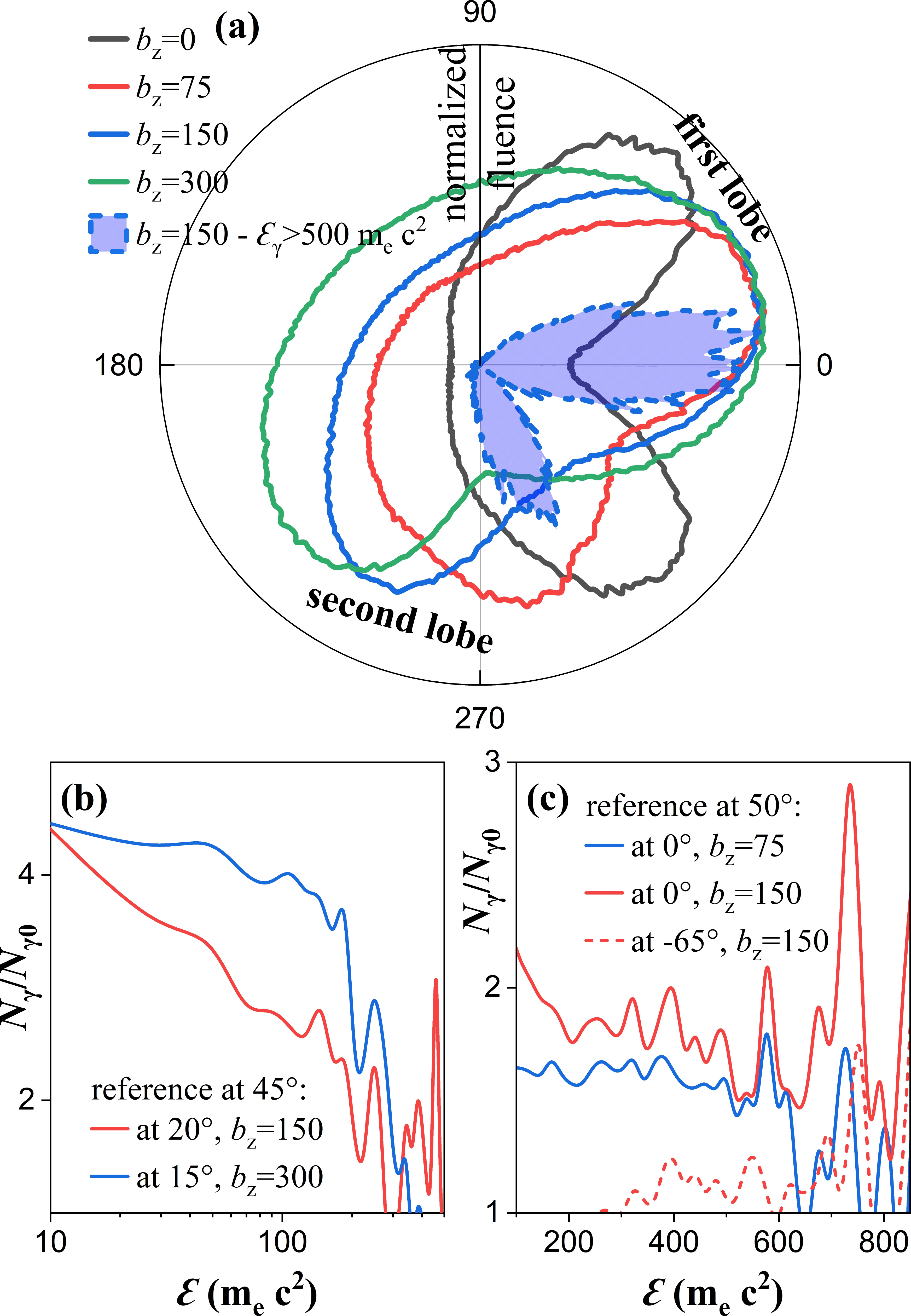}
  \caption{(a) Normalized fluence in radial directions on the xy-plane, within $1 \degree$ full angle. The cases $b=0$, $b_z=75$, $b_z=150$, $b_z=300$ correspond to the black, red, blue and green lines respectively. The blue filled area corresponds to the $b_z=150$ case, by considering only \textgamma-photons of energy larger than $500 m_e c^2$. (b) The ratio of $b_z$ to $b=0$ \textgamma-photon energy spectra, measured at peak $I_\Omega$ within an $1 \degree$ full angle. (c) The ratio of $b_z$ to $b=0$ \textgamma-photon energy spectra, measured at peak $I_\Omega$ (where here for $I_\Omega$ we consider only \textgamma-photons of energy larger than $500 m_e c^2$), within a $10 \degree$ full angle. The $b_z \approx 150$ and $b_z \approx 300$ cases correspond to the red and blue lines respectively.}
  \label{fig:5}
\end{figure} % figure figure figure figure figure figure figure figure figure figure figure

\par The \textgamma-photon energy spectrum depends on the \textgamma-photon detection angle. This is seen in Fig. \ref{fig:5}(a), where the normalized fluence is shown, for the three $b_z$ cases examined as labeled in the figure. For the $b=0$ case two symmetric lobes are obtained. However, by increasing the CMF value the first lobe shifts closer to the laser propagation axis while the second lobe shifts far from the axis. In addition, the figure depicts with the blue shadowed region the case where only \textgamma-photons of energy larger than $500 m_e c^2$ are considered. There, a dominant lobe exists on the laser propagation axis with a divergence of approximately $10 \degree$ at full-width-at-half-maximum, and another less prominent lobe at angles of $-45 \degree$, $-65 \degree$ and $-110 \degree$ for the $b_z=75$, $b_z=150$ and $b_z=300$ cases respectively.

\par For applications, one needs to know how the \textgamma-photon number changes per energy interval. Therefore, we calculate the ratio of the \textgamma-photon spectrum at the peak $I_\Omega$ location for the $b_z$ case, to the reference case. For $b_z=150$, $I_\Omega$ peaks at $20 \degree$ to the laser propagation axis, and for $b_z=300$ it peaks at $15 \degree$; for $b=0$, the reference $I_\Omega$ peaks at $45 \degree$. The spectra ratios for the $b_z=150$ and $b_z=300$ cases are shown in Fig. \ref{fig:5}(b) with red and blue lines respectively. In both cases, the \textgamma-photon number, $N_\gamma$, is increased three (for higher \textgamma-photon energies) to five (for lower \textgamma-photon energies) times compared to the reference \textgamma-photon number, $N_{\gamma0}$. However, considering peak $I_\Omega$ only of higher energy \textgamma-photons, results in spectra ratios as shown in Fig. \ref{fig:5}(c). The lobe on the laser propagation axis contains approximately twice as high energy \textgamma-photons compared to the reference case, taken at $50 \degree$. The second lobe contains approximately same number of high energy \textgamma-photons as in the reference case.

%---------------------------------------------------------------------------------------------------------------------------------------------------------------------------

%{\color{blue}

\par In conclusion, the spatial and spectral distributions of the emitted \textgamma-photons are obtained via PIC simulations, where an $a=350$ laser interacts with a lithium foil. The radiation reaction force alters the trajectory of a single-electrons moving under the influence of an ultraintense laser in addition to a CMF. The radiation reaction force obtains high values when the CMF is transverse to the laser propagation axis. If the CMF aligns with the laser electric field then the \textgamma-photons are emitted in two broad lobes, whilst if it coincides with the laser magnetic field then the electron moves strictly on the xy-plane and the \textgamma-photons are emitted mostly along that plane. Specifically, the radiant intensity for the $b_z=300$ case is increased by a factor of more than three. Highest $\kappa_\gamma$ is obtained for the $b_z=150$ case, reaching up to $60 \kern0.2em \%$, three times higher than the reference case. Moreover, the amplitude of the lower and higher part of the \textgamma-photon energy spectrum is increased by a factor of five and two respectively. The enhancement of the low energy \textgamma-photon number at optimal emission angles suits photonuclear reactions \cite{2022_KolenatyD}. The higher energy \textgamma-photon number is doubled along the laser propagation axis, necessary for electron-positron pair generation the nonlinear Breit-Wheeler process \cite{2023_MacLeodAJ}.

%}

%%%%%%%%%%%%%%%%%%%%%%%%%%%%%%%%%%%%%%%%%%%%%%%%%%%%%%%%%%%%%%%%%%%%%%
%%%%%%%%%%%%%%%%%%%%%%%%%%%%%%%%%%%%%%%%%%%%%%%%%%%%%%%%%%%%%%%%%%%%%%

%%%%%%%%%%%%%%%%%%%%%%%%%%%%%%%%%%%%%%%%%%%%%%%%%%%%%%%%%%%%%%%%%%%%%%
%%%%%%%%%%%%%%%%%%%%%%%%%%%%%%%%%%%%%%%%%%%%%%%%%%%%%%%%%%%%%%%%%%%%%%

\par This work was supported by the Ministry of Education, Youth and Sports of the Czech Republic through the e-INFRA CZ (ID:90254). The EPOCH code is in part funded by the UK EPSRC grants EP/G054950/1, EP/G056803/1, EP/G055165/1 and EP/M022463/1.

%\par This work is supported by the project “Advanced research using high intensity laser produced photons and particles” (ADONIS) (CZ.02.1.01/0.0/0.0/16 019/0000789) from the European Regional Development Fund. The EPOCH code is in part funded by the UK EPSRC grants EP/G054950/1, EP/G056803/1, EP/G055165/1 and EP/M022463/1.

\bibliography{biblio} % Produces the bibliography via BibTeX.

\end{document}